\documentclass[usenatbib]{mnras}

\usepackage{newtxtext,newtxmath}

\usepackage[T1]{fontenc}
\usepackage{ae,aecompl}

\usepackage{graphicx}	
\graphicspath{{Images/}}
\usepackage{amsmath}	
\usepackage{amssymb}	
\usepackage[table,xcdraw]{xcolor}    
\usepackage{subfig}
\usepackage{comment}
\usepackage[normalem]{ulem}

\definecolor{royalazure}{rgb}{0.0, 0.22, 0.66}
\definecolor{auburn}{rgb}{0.43, 0.21, 0.1}
\definecolor{bostonuniversityred}{rgb}{0.8, 0.0, 0.0}

\usepackage{hyperref}
\hypersetup{
    colorlinks=true,
    linkcolor=bostonuniversityred,
    linktoc=blue,
    filecolor=blue,      
    urlcolor=auburn,
    citecolor=royalazure
}

\title[Survival of planets in self-gravitating discs]{Planet Migration in Self-Gravitating Discs: Survival of Planets}

\author[S. Rowther \& F. Meru]{
Sahl Rowther,$^{1,2}$\thanks{sahl.rowther@warwick.ac.uk}
Farzana Meru$^{1,2}$
\\
$^{1}$Centre for Exoplanets and Habitability, University of Warwick, Coventry CV4 7AL, UK\\
$^{2}$ Department of Physics, University of Warwick, Coventry CV4 7AL, UK\\
}

\date{Accepted 2020 June 1. Received 2020 May 18; in original form 2020 April 6}

\pubyear{2020}

\begin{document}
\label{firstpage}
\pagerange{\pageref{firstpage}--\pageref{lastpage}}
\maketitle

\begin{abstract}
We carry out three-dimensional SPH simulations to study whether planets can survive in self-gravitating protoplanetary discs. The discs modelled here use a cooling prescription that mimics a real disc which is only gravitationally unstable in the outer regions. We do this by modelling the cooling using a simplified method such that the cooling time in the outer parts of the disc is shorter than in the inner regions, as expected in real discs.  We find that both giant ($> M_{\mathrm{Sat}}$) and low mass ($< M_{\mathrm{Nep}}$) planets  initially migrate inwards very rapidly, but are able to slow down in the inner gravitationally stable regions of the disc  {without needing to open up a gap}. This is in contrast to previous studies where the cooling was modelled in a more simplified manner where regardless of mass, the planets were unable to slow down their inward migration. This shows the important effect the thermodynamics has on planet migration. In a broader context, these results show that planets that form in the early stages of the discs' evolution, when they are still quite massive and self-gravitating, can survive.
\end{abstract}

\begin{keywords}
hydrodynamics -- protoplanetary discs -- planet-disc interactions
\end{keywords}



\section{Introduction}

During the last couple of decades the number of exoplanets discovered has increased drastically thanks to advances in technology. As of current writing 4154 exoplanets have been confirmed according to the \href{https://exoplanetarchive.ipac.caltech.edu/}{NASA Exoplanet Archive}\footnote{https://exoplanetarchive.ipac.caltech.edu/}, and this number will continue to increase significantly with ongoing missions such as TESS and future missions such as PLATO

Over the last few years, advances in imaging have resulted in surveys of high-resolution images of young protoplanetary discs. A large number of these discs have substructures, such as rings and gaps \citep{2015ALMA,2016Andrews,2018Andrews,2018Huang,2018Dipierro,2020Booth}. A common explanation for their presence is planets \citep{2015Dipierro,2016Dipierro,2018Long,2018Zhang,2018Dullemond}. However, many of these discs appear to be quite young, which would require giant planets to form very early in the lifetime of the disc. \cite{2017Ansdell} also find that even in the younger discs, there it not enough material to form the core of a giant planet. This suggests that if planets are the source of gaps and rings, then planet formation must occur at a much faster timescale. In the early stages of disc evolution, the disc is expected to be quite massive and large in size. However, there is a growing body of evidence that such discs can exist \citep{2016Tobin, 2016Perez}.

An issue with this model is whether planets that form early on in these discs, when the discs are potentially gravitationally unstable, can survive. Previous studies such as \cite{2011Baruteau} have found that giant planets in self-gravitating discs rapidly migrate inwards with no signs of gap opening. It is much harder to open up gaps in self-gravitating region of a disc due to the high levels of turbulence. \cite{2015Malik} also found it difficult to open up gaps,  even with $30M_{\mathrm{Jup}}$ companions. However, these studies used a simple dimensionless globally constant cooling parameter $\beta$, to cool the disc, {where $\beta$ is defined as the ratio between the cooling time and the local orbital time. Although this method is quick as it gives a simple expression for the cooling time, it does have its issues}. The main one being that as the disc evolves, spiral structure is formed throughout the disc. This behaviour is not expected from realistic discs where they should only be gravitationally unstable in the outer regions \citep{2005Rafikov,2009Stamatellos, 2009Rice, 2009Clarke}.

Other studies such as \cite{2018Stamatellos, 2015Stamatellos} have found that when realistic thermodynamics are included, giant planets can open up gaps and slow their inward migration. However, in their simulations the planets were also allowed to accrete material which could aid gap opening. It is therefore unclear if their planet migration slowed down due to the mass or once they reached the gravitationally stable inner disc. The gravitational instability of a disc can be defined by $Q$, the Toomre parameter \citep{1964Toomre}
\begin{equation}
Q = \frac{c_{s}\Omega}{\pi G \Sigma},
\end{equation}
where $c_{s}$ is the sound speed of the disc, $\Sigma$ is the disc surface density, $\Omega$ is the angular frequency, and $G$ is the gravitational constant. 

\cite{2009Cossins} showed that the amplitude of density fluctuations decreases at a rate inversely proportional to $\beta$. This means that in constant $\beta$ models as in \cite{2011Baruteau} and \cite{2015Malik}, there will be density fluctuations triggered by gravitoturbulence throughout the disc. These fluctuations can result in mass deficits around the planet and induce stochastic kicks (also seen in discs with MHD turbulence; \citealt{2004Nelson}) or even Type III migration. 

The motivation behind this work is to determine whether planets  -- irrespective of how they form -- can slow down their migration in the inner gravitationally stable part of the disc. To accomplish this, we mimic the effects of more realistic thermodynamics by using the aforementioned simple cooling method but we vary it with radius such that the cooling timescale in the outer regions is much smaller than the inner regions so that the disc is only gravitationally unstable in the outer part.  A larger $\beta$ {(and hence a longer cooling timescale)} in the inner gravitationally stable region, resulting in smaller density fluctuations could provide a safe haven for the planet. The main advantage of this method is to avoid having to use the more complicated and computationally expensive method of radiative transfer. We study the migration of planets in these more realistic discs to determine whether they can slow their inward migration and survive.

The outline of this paper is as follows. In Section \ref{sec:methods}, we describe how the simulations are set up and how the disc thermodynamics is treated. In Section \ref{sec:res}, we present the comparison of the new implementation of the cooling parameter with the standard implementation. We compare the effect it has on the evolution of the disc, and also how planet migration is affected. In Section \ref{sec:num}, we investigate the impact of numerics on the results. In Sections \ref{sec:disc}, we put the work done here in context of past work and observations. Finally, we make conclusions in Section \ref{sec:conc}.

\section{Model}
\label{sec:methods}
\subsection{Numerical Method}

We perform 3D hyrdrodynamic simulations using \texttt{PHANTOM}, a Smoothed Particle Hydrodynamics (SPH) code developed by \cite{2018Price}. Most of the simulations presented here are run with 2 million gas particles. To ensure the results are not affected by resolution, a subset of the simulations are also done with 1 and 4 million particles. The star and planets are modelled as sink particles \citep{1995Bate}. The accretion radius of the star is set to be equal to the disc inner boundary, $R_{in}$. The accretion radius of each planet is set to 0.001, which is much smaller than their Hill radius. This is to prevent the planets from accreting too much material in order to perform as close a comparison to \cite{2011Baruteau} and \cite{2015Malik}.

To model shocks, we use an artificial viscosity switch that utilises the time derivative of the velocity divergence introduced by \cite{2010Cullen}. The artificial viscosity parameter $\alpha_{v}$, is varied depending on how far away the shock is. Close to the shock, it is a maximum, $\alpha_{\text{max}} = 1$. Far from the shocks, it is a minimum, $\alpha_{\text{min}} = 0$. The artificial viscosity coefficient $\beta_{v}$ is set to 2.

\subsection{Initial Conditions}
\label{sec:IC} 

We model a disc between $R_{in} = 1$ and $R_{out} = 25$, with disc to star mass ratio of 0.1. It should be noted that these simulations are scale free. The initial surface mass density is set as a smoothed power law and is given by
\begin{equation}
\Sigma = \Sigma_{0}  \left ( \frac{R}{R_{0}} \right)^{-1} f_{s},
\end{equation}
where $\Sigma_{0}$ is the surface mass density at $R=R_{0}=1 $ and ${f_{s} = 1-\sqrt{R_{in}/R}}$ is {the factor used to smooth the surface density at the inner boundary of the disc}.

The energy equation is given by
\begin{equation}
\frac{\mathrm{d}u}{\mathrm{d}t} = -\frac{P}{\rho} \left ( \nabla \cdot \mathbf{v} \right) + \Lambda_{\mathrm{shock}} - \frac{\Lambda_{\mathrm{cool}}}{\rho}
\end{equation}
where $u$ is the specific internal energy, the first term on the RHS is the $P\mathrm{d}V$ work, $\Lambda_{\mathrm{shock}}$ is a heating term that is due to the artificial viscosity used to correctly deal with shock fronts, and $\Lambda_{\mathrm{cool}}$ is a cooling term to cool the disc. 

We assume an adiabatic equation of state. For an ideal gas this relates the pressure $P$, to the density $\rho$, and specific internal energy by
\begin{equation}
\label{EqOfState}
P = (\gamma - 1) \rho u = \frac{c_{s}^{2} \rho}{\gamma},
\end{equation}
where the adiabatic index is $\gamma = 5/3$ and the sound speed is ${c_{s} = \sqrt{\gamma k_{\mathrm{B}}T/\mu m_{\mathrm{H}}}}$. The mean molecular weight is assumed to be $\mu=2.381$. The initial temperature profile is expressed using a power law as
\begin{equation}
T = T_{0} \left ( \frac{R}{R_{0}} \right)^{-0.5}.
\end{equation}
The disc aspect ratio $H/R$, is set to 0.05 at $R=R_{0}$.
\subsection{The cooling timescales}

In this work, we simply set the cooling term as
\begin{equation}
\Lambda_{\mathrm{cool}} = \frac{\rho u}{t_{\mathrm{cool}}}
\end{equation}
where the cooling time is given by
\begin{equation}
t_{\mathrm{cool}} = {\beta}{\Omega^{-1}}.
\end{equation}
Here, $\beta$ is the cooling parameter. Under the assumption that the transfer of angular momentum driven by gravitoturbulence occurs locally, we can relate the cooling parameter $\beta$ with the $\alpha$ parameter  by \citep{2001Gammie}
\begin{equation}
\label{betaalpha}
\alpha = \frac{4}{9}\frac{1}{\gamma(\gamma - 1)} \frac{1}{\beta}.
\end{equation}

\begin{figure}
\begin{center}

\includegraphics[width= \linewidth]{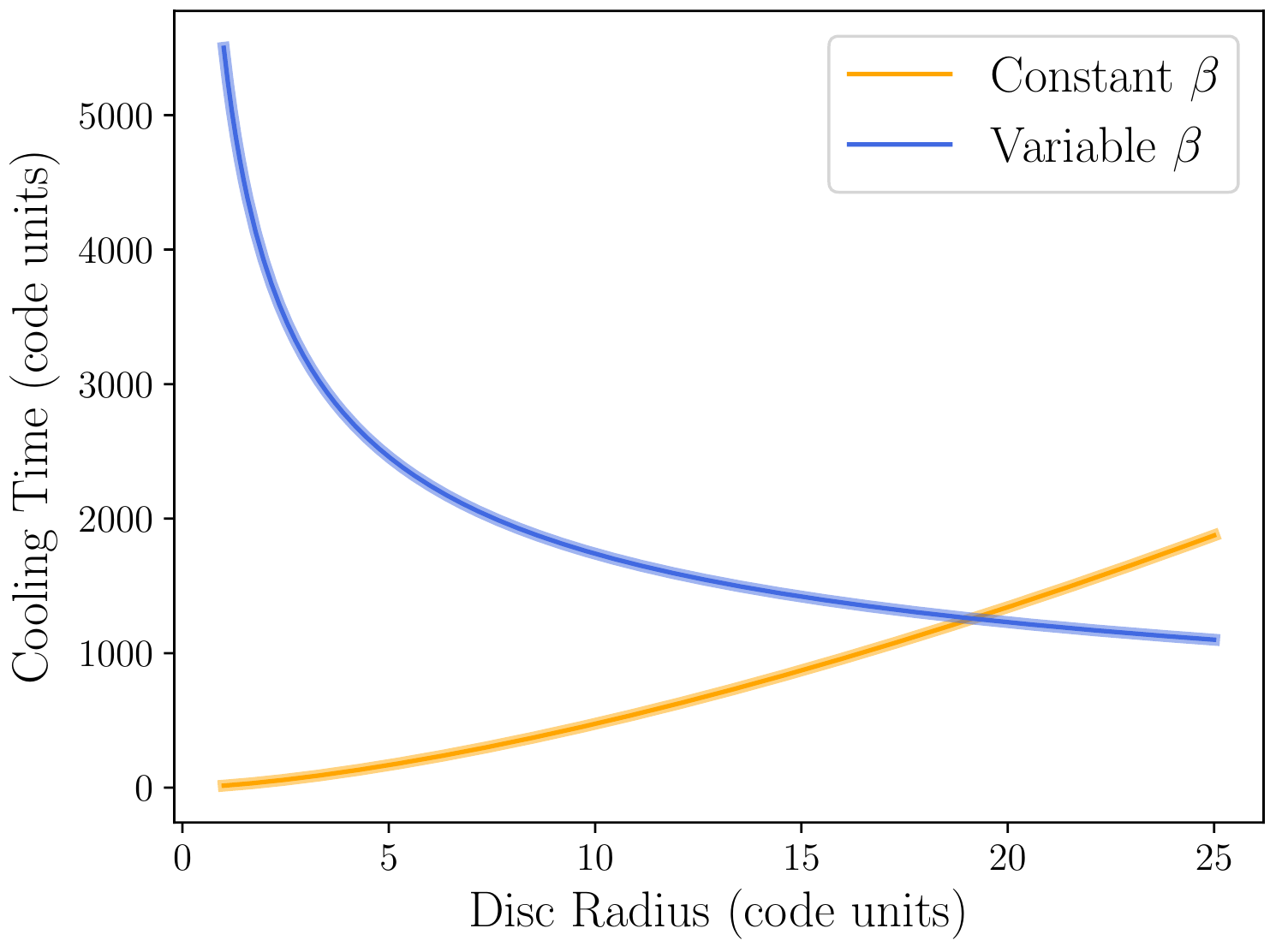}
\caption[]{Comparison between a constant ($\beta = 15$) and a variable (${\beta = \beta_{0}(R/R_{0})^{-2}}$) cooling parameter. The cooling time in the inner regions is significantly longer using a variable $\beta$.}
\label{CoolingTime}
\end{center}
\end{figure}

We perform two sets of simulations with different cooling implementations to model the disc thermodynamics. {Firstly,} we define $\beta$ as
\begin{equation}
\label{eq:VB}
\beta = \beta_{0} \left ( \frac{R}{R_{0}} \right)^{-2},
\end{equation}
where $\beta_{0} = 5500$. Since $\beta$ varies with radius, so does $\alpha$. In these discs $\alpha$ ranges from $\sim 10^{-4}$ in the inner regions to $\sim10^{-2}$ in the outer regions. {Secondly,} to compare with previous studies, we run a set of simulations where the cooling parameter is globally constant and set to $\beta = 15$. Using equation \ref{betaalpha}, this is equivalent to $\alpha = 0.027$. This value and the normalisation constant in equation \ref{eq:VB} is chosen to ensure that while the disc cools fast enough to develop spiral structure, it does not fragment into clumps. This is done so that the migration of the planet is not  affected by the interaction with the clumps.
The difference between these cooling timescales on disc evolution and planet migration is shown in Figure \ref{CoolingTime} where it can be seen that using the variable $\beta$ defined in equation \ref{eq:VB} results in the cooling timescale in the inner regions being significantly longer than with constant $\beta$. Hence, the inner regions now cool slowly enough to avoid becoming gravitationally unstable, while the outer regions still cool fast enough to become gravitationally unstable and form spiral structure.
 
\subsection{Embedding the planets}

After the disc has evolved long enough to form spiral structure (around 10 orbits at the outer edge, $R=25$), a planet is added at $R=20$. To explore the effect the variable cooling parameter has on {planet} migration, the evolution of both giant and low mass planets is followed. 
{The giant planets have planet to star mass ratios of ${q = 2.857 \times 10^{-4}}, \  9.543 \times 10^{-4}$ and $9.543 \times 10^{-3}$. The low mass planets have planet to star mass ratios of  ${q = 3 \times 10^{-6}}, \  {3 \times 10^{-5}}$ and $5.149 \times 10^{-5}$. This is equivalent to a $1M_{\mathrm{Sat}}$, $1M_{\mathrm{Jup}}$, and a $5M_{\mathrm{Jup}}$ planet and a $1M_{\oplus}$, $10M_{\oplus}$, and a $1M_{\mathrm{Nep}}$ planet in a $0.1M{\odot}$ disc around a $1M_{\odot}$ star.}

\section{Results}
\label{sec:res}

The simulations presented here are done in two steps. First the disc is set up as described in Section \ref{sec:IC} and is allowed to evolve until {spiral structure develops}. The evolution of the disc is described in Section \ref{sec:DE} Then after the inclusion of planets of various masses, the simulations are resumed and the planet's migration is shown in Section \ref{sec:PM}. These are done with both a constant and variable $\beta$.

\subsection{Disc evolution}
\label{sec:DE}

\begin{figure*}
\begin{center}
\includegraphics[width= 0.775\textwidth]{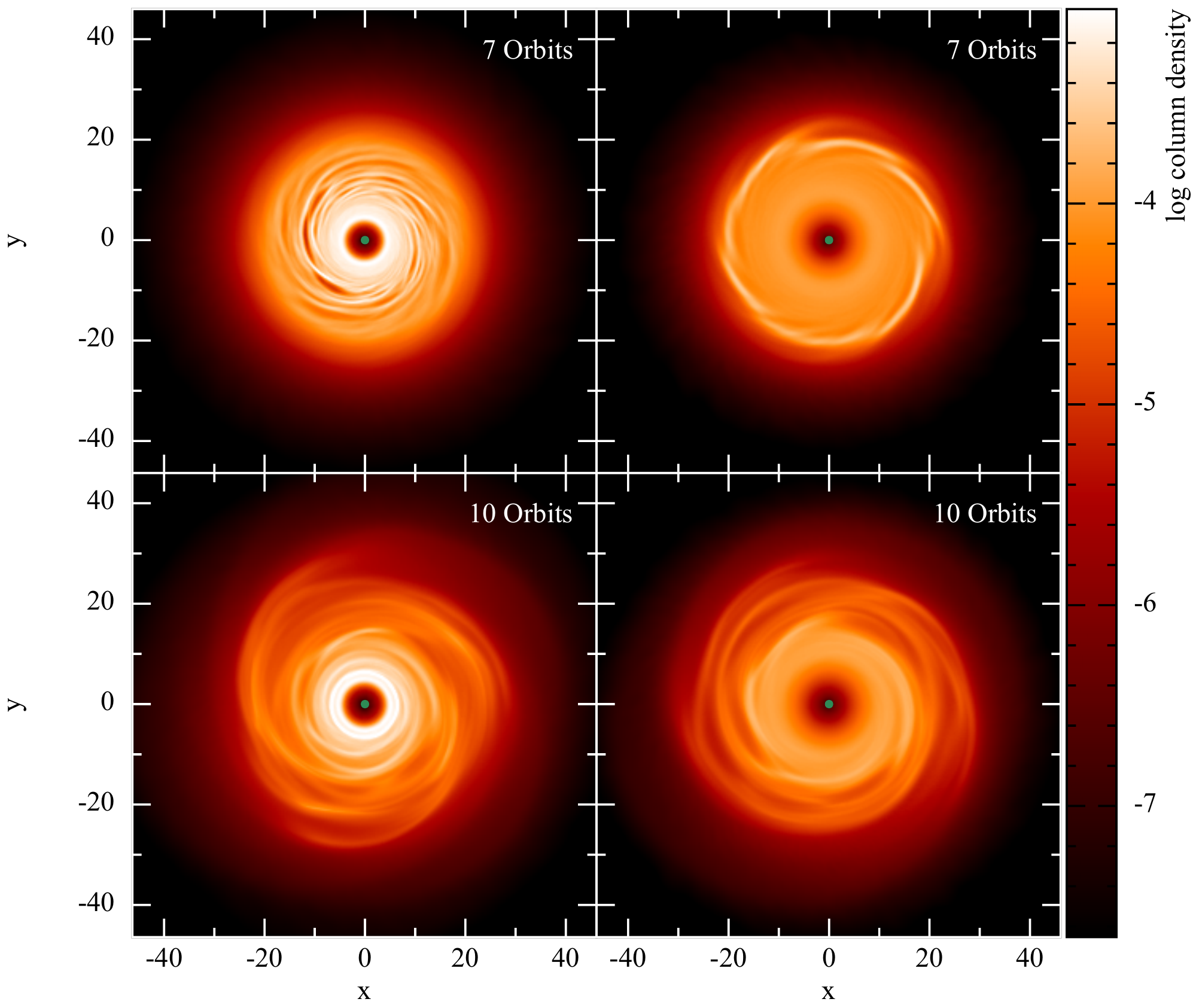}
\caption[]{{Comparison of the evolution of two discs evolved with a constant cooling parameter $\beta$ given by equation \ref{eq:VB} (left), and a variable $\beta$ (right), after 7 and 10 orbits (at $R=25$)}. The top row shows the two discs at an early stage where the spiral structure has just begun to form. The bottom row shows both discs at a later stage just before a planet is added. This figure highlights that a variable $\beta$ mimics a more realistic disc where the disc is only gravitationally unstable in the outer regions.}
\label{fig:discCompare}
\end{center}
\end{figure*}

\begin{figure*}
\begin{center}
\centering\captionsetup[subfloat]{labelfont=bf}
\subfloat[Comparison of the disc temperature.]{\includegraphics[width = 0.48\textwidth]{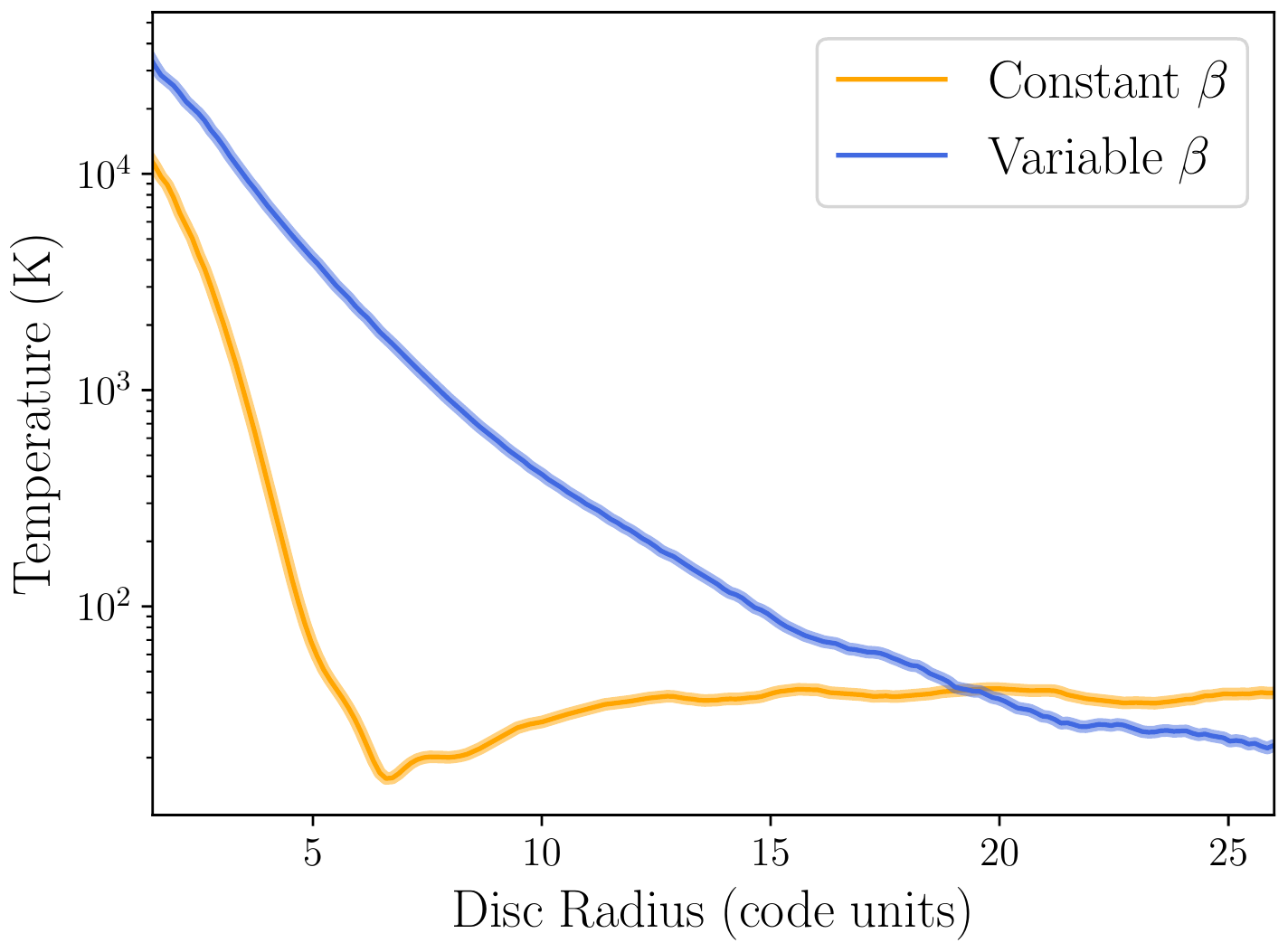}\label{Tdiscs}} 
\hspace{0.5cm}
\subfloat[Comparison of the Toomre parameter, $Q$.]{\includegraphics[width = 0.48\textwidth]{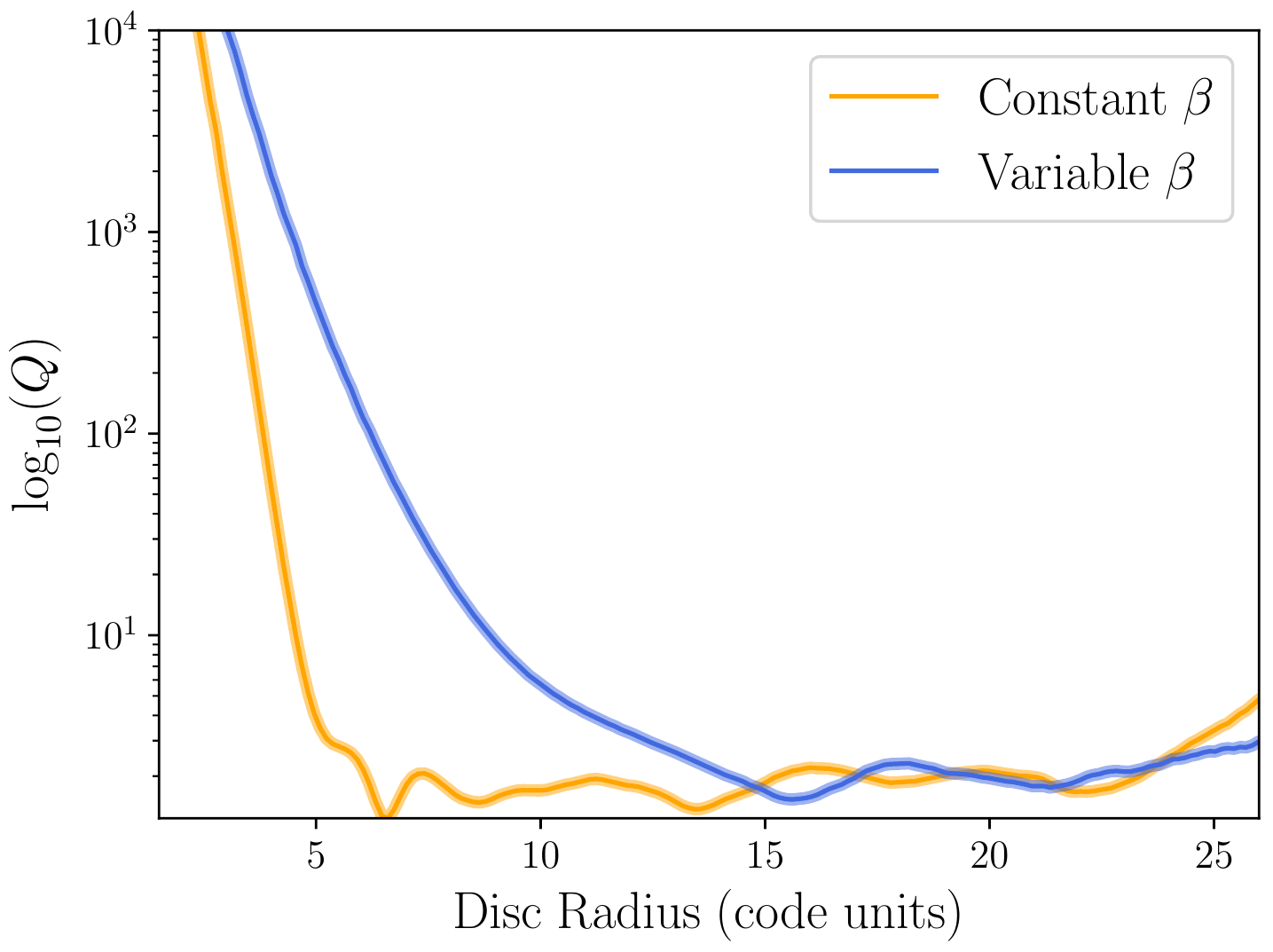}\label{Qdiscs}} 
\caption[]{Azimuthally-averaged temperature (left) and Toomre parameter (right) for the variable and constant $\beta$ simulations after 10 orbits at $R=25$. This figure shows why the disc cooled with a variable $\beta$ is only gravitationally unstable in the outer regions. Due to the significantly longer cooling times in the inner regions, the inner disc cannot cool fast enough, which results in the  Toomre parameter not reaching the required values for gravitational instability. {Although the temperatures are quite high, it should be noted that if these discs were scaled up to sizes of typical gravitationally unstable discs, while scaling the initial $Q$ profile to be identical, the temperatures would be more sensible.} The discs at this time are shown in the bottom row in Figure \ref{fig:discCompare}.}
\label{fig:discTempCompare}
\end{center}
\end{figure*}

{Both discs start with identical surface density and temperature profiles, so that the evolution of the disc is determined by how it cools}. The difference is seen in Figures \ref{fig:discCompare} and \ref{fig:discTempCompare}. In Fig \ref{fig:discCompare}, the density rendered plots of both discs is shown at two different times. The top row shows the disc early in its evolution just as the spiral structure begins to form. The disc that evolves with  constant $\beta$  develops spiral structure from the inside  first (top left) and develops spiral structure throught the disc (bottom left), which is consistent with past studies. Whereas the disc that  evolves with variable $\beta$ only develops spiral structure in the outermost regions  (right). Similarly, the bottom row shows how the discs differ shortly before a planet is added. This behaviour in the bottom right panel is more in line with what is expected from realistic discs, where discs are expected to be gravitationally unstable only in the outer regions  \citep{2005Rafikov, 2009Stamatellos, 2009Rice, 2009Clarke}.

The difference between the two discs is due to the different expressions for $\beta$ used. Since the variable $\beta$ increases with decreasing radius, it results in the cooling time in the inner regions being much larger. 
This effect of the cooling time on the disc evolution is shown in Figure \ref{fig:discTempCompare} by comparing the temperature and the the Toomre parameter of both discs at the same moment in time. Using equation \ref{EqOfState} and the expression for the sound speed, the temperature as the disc evolves can be calculated as
\begin{equation}
T = \frac{\mu m_{\mathrm{p}}}{k_{\mathrm{B}}} \left ( \gamma -1 \right) u.
\end{equation}

{A consequence of the variable $\beta$ model is the disc cools extremely inefficiently in the innermost regions of the disc, resulting in very high temperatures as seen in Figure \ref{Tdiscs}. However, the simulations present here are scale free. Hence, although the temperatures are quite high, they would be more in line with expected values if the disc was scaled up to sizes of typical gravitationally unstable disc, while also scaling the initial $Q$ profile to be identical. Furthermore, the high temperatures at $R<10$ is not a major concern for the purposes of this work since the planet reaches this part of the disc.}

Figure \ref{Qdiscs}  compares the azimuthally averaged  Toomre parameter, $Q$.
Since $Q$ is dependent on the temperature via the sound speed, this results in $Q \gg 1.7$ in the inner regions. Hence, the conditions for spiral structure to form is only satisfied in the outer parts \citep{2007Durisen}.

\subsection{Planet migration}
\label{sec:PM}

\subsubsection{Giant Planets}
\label{GP}

\begin{figure*}
\begin{center}
\includegraphics[width= \linewidth]{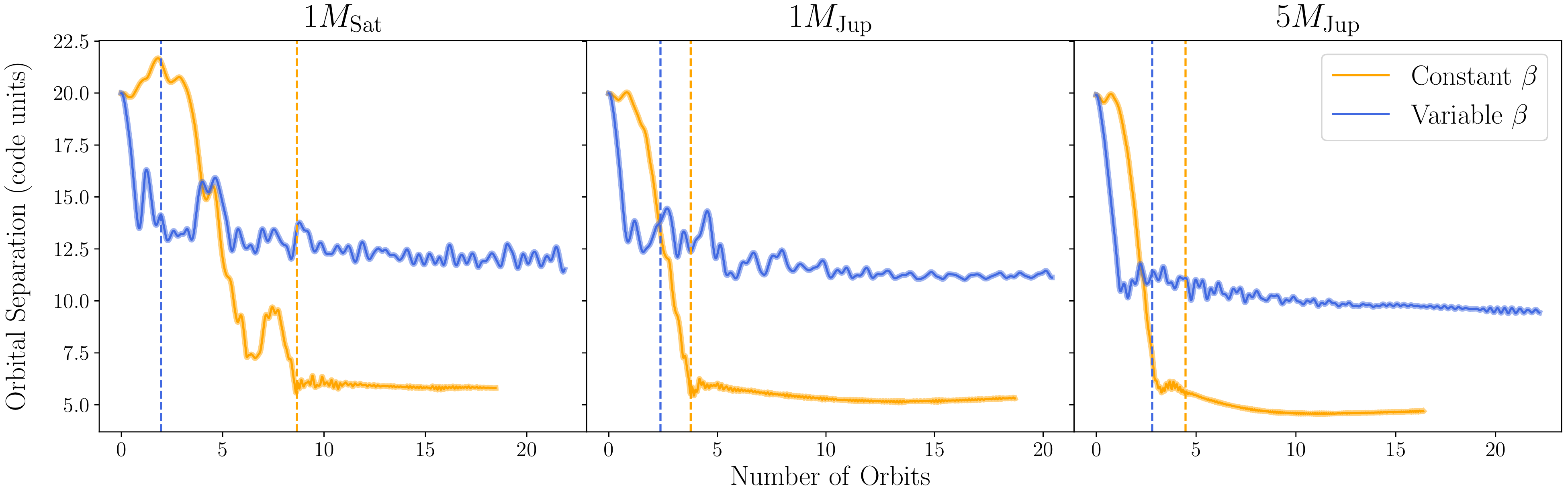}
\caption[]{Migration tracks of a $1M_{\mathrm{Sat}}$ (left), $1M_{\mathrm{Jup}}$ (middle), and a $5M_{\mathrm{Jup}}$ (right) planet in a disc cooled with a variable $\beta$ (blue) and constant $\beta$ (orange). {The planet slows down in the inner disc modelled with a variable $\beta$}, whereas with a constant $\beta$, the inward migration does not slow down and the planets reach the inner edge of the disc. The dashed lines are defined as the time when migration has slowed down (or reached the inner edge).}
\label{migration}
\end{center}
\end{figure*}

\begin{figure*}
\begin{center}
\includegraphics[width= 0.995\linewidth]{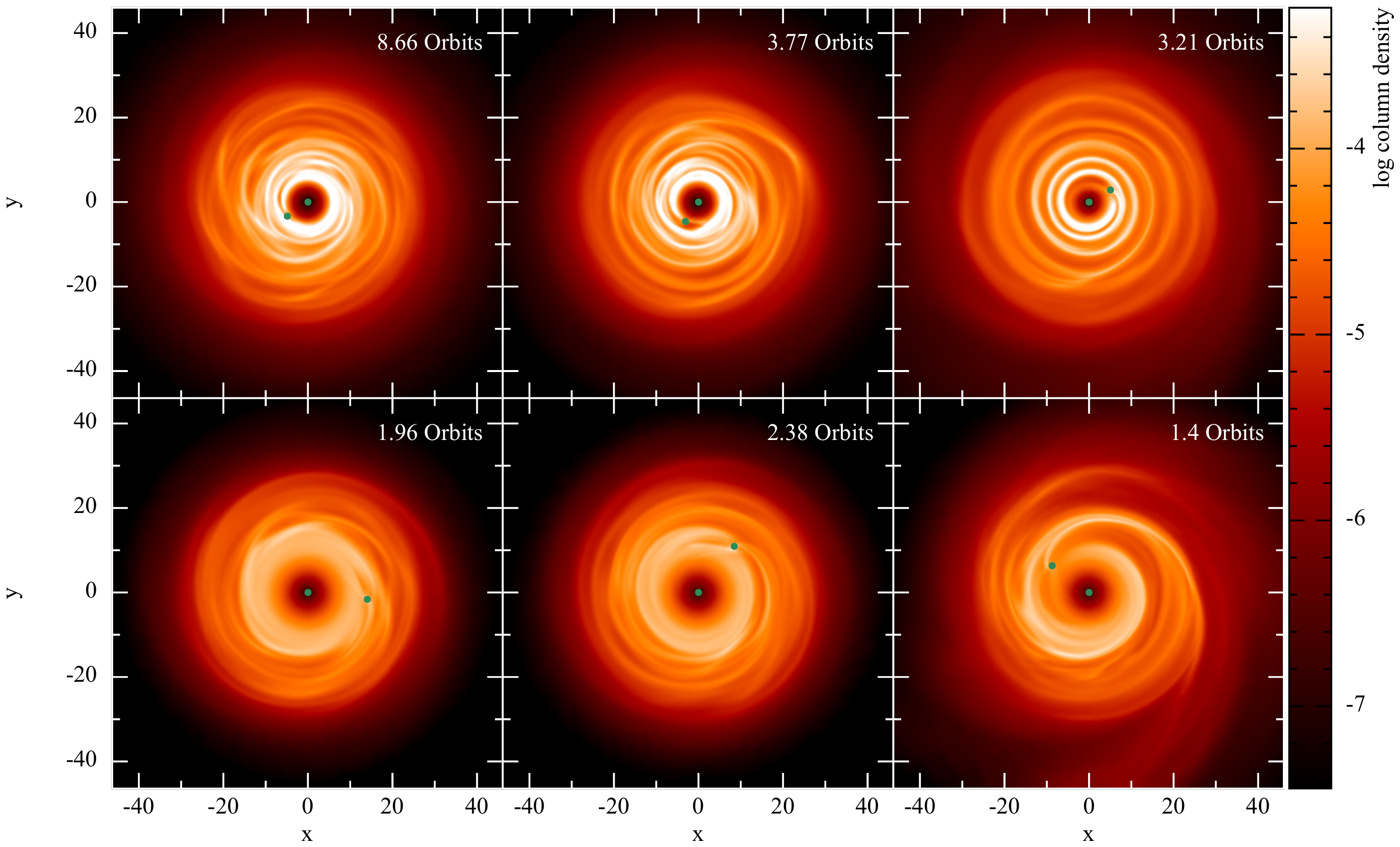}
\caption[]{This figure shows the location of a $1M_{\mathrm{Sat}}$ (left), a $1M_{\mathrm{Jup}}$ (middle), and a $5M_{\mathrm{Jup}}$ planet (right) soon after their migration has slowed down. This is shown for a disc modelled with a constant $\beta$ (top row) and a variable $\beta$ (bottom row). The planets in a constant $\beta$ disc only stop migrating when they reach the inner edge of the disc. Whereas, with a variable $\beta$, migration slows down before reaching the disc's inner edge. The times in each panel correspond to the dashed lines in Figure \ref{migration}.}
\label{migrationPlot}
\end{center}
\end{figure*}

To investigate the effect the variable $\beta$ has on planet migration, a single planet is added at $R=20$ in the simulations in Section \ref{sec:DE} for both the constant and variable $\beta$ discs. The value of $\beta$ at the initial location of the planet is roughly similar in both discs (see Figure \ref{CoolingTime}).

\begin{figure}
\begin{center}
\includegraphics[width= 1.0\linewidth]{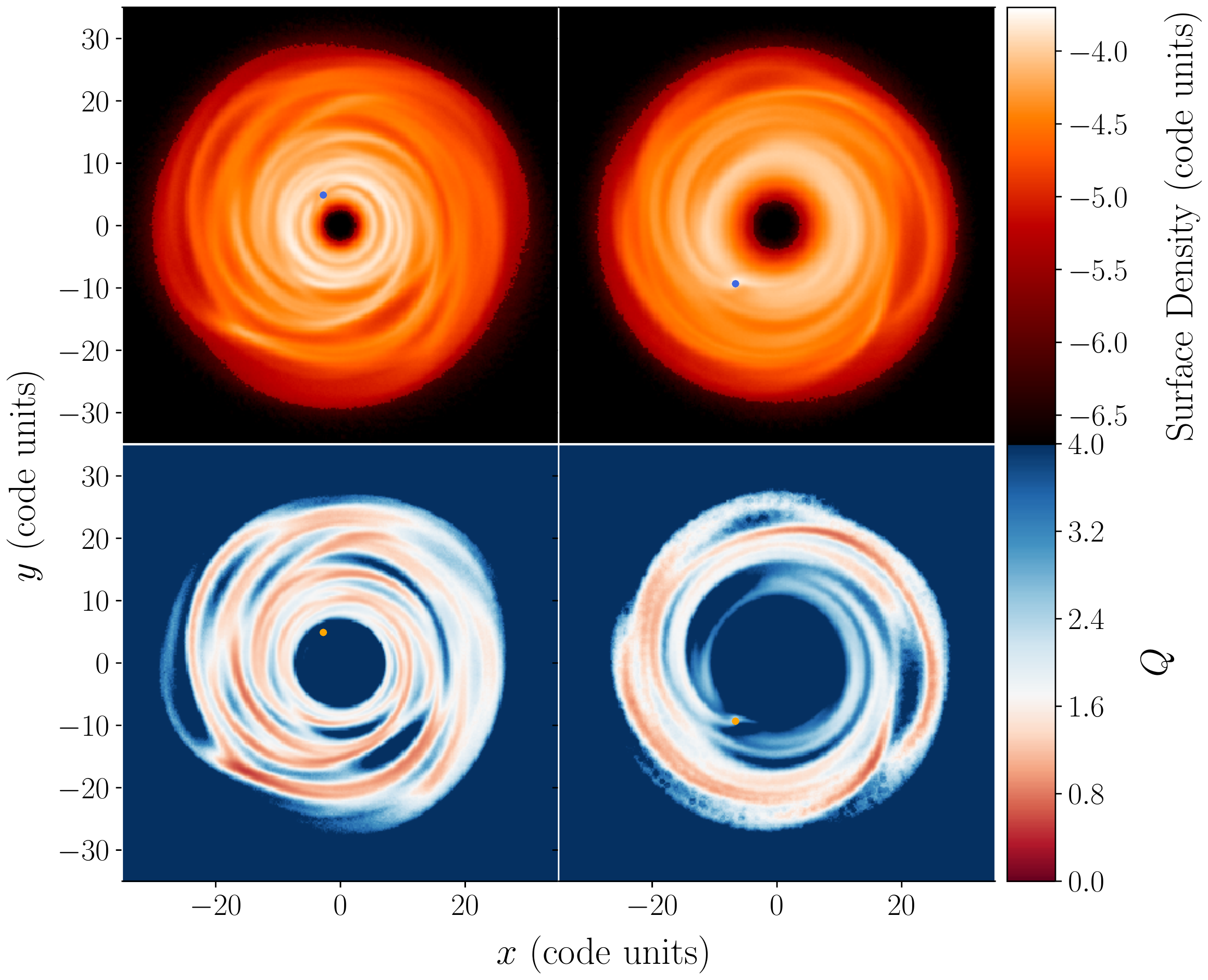}  
\caption[]{Surface density (top) and 2D Toomre parameter (bottom) plots of  the disc soon after a $1M_{\mathrm{Jup}}$ planet reaches the inner edge of a constant $\beta$ (left) and {slows its migration} in a variable $\beta$ (right) disc.  The 2D $Q$ plots show how gravitationally unstable the disc is. It can be seen that the migration of a $1M_{\mathrm{Jup}}$ planet slows down in the gravitationally stable part of the disc.}
\label{2DQ}
\end{center}
\end{figure}

To compare with previous studies that also used $\beta$-cooling to study planet migration in self-gravitating discs, we first evolve the planets in a constant $\beta$ disc. From the migration tracks, shown in orange in Figure \ref{migration}, and in the top row of Figure \ref{migrationPlot}, it can be seen that the planets reach the inner edge of the disc in a few orbits. These compare well with previous studies such as \cite{2011Baruteau}. However, in a disc cooled with a variable $\beta$, the migration tracks are no longer the same. Although both initially migrate inwards very rapidly, with a variable $\beta$, {the inward migration slows down in the inner gravitationally stable part of the disc} as seen in the bottom row of Figure \ref{migrationPlot}.  {This is more clearly seen in Figure \ref{2DQ} which shows the surface density (top row) and a 2D map of the Toomre stability parameter $Q$ (bottom row) of a constant $\beta$ disc (left column) when a $1M_{\mathrm{Jup}}$ planet reaches the disc's inner edge, and of a variable $\beta $ disc (right column) when the planet's migration slows down. It can be seen that unlike in a constant $\beta$ disc, the inner region of a variable $\beta$ disc is gravitationally stable, and it is in this region where the planets are able to slow down their inward migration.
}

\subsubsection{Stochastic kicks}

As seen from Figure \ref{migration}, planet migration in self-gravitating discs is not a smooth process. The planets can experience random kicks that push them either inwards or outwards. To ensure that these random kicks due to the turbulent nature {of the gravitationally unstable disc} do not impact the results of the giant planet simulations, {the simulations in Section \ref{GP} are repeated another 3 times}. In each new run the planets are embedded at the same distance, {but started 0.2 orbits earlier}. Starting at a different time means that the turbulent structure around the planet is different {in each simulation}. Hence, it is a good test to determine whether the turbulence has a significant impact on planet migration. 

\begin{figure}
\begin{center}
\includegraphics[width= \linewidth]{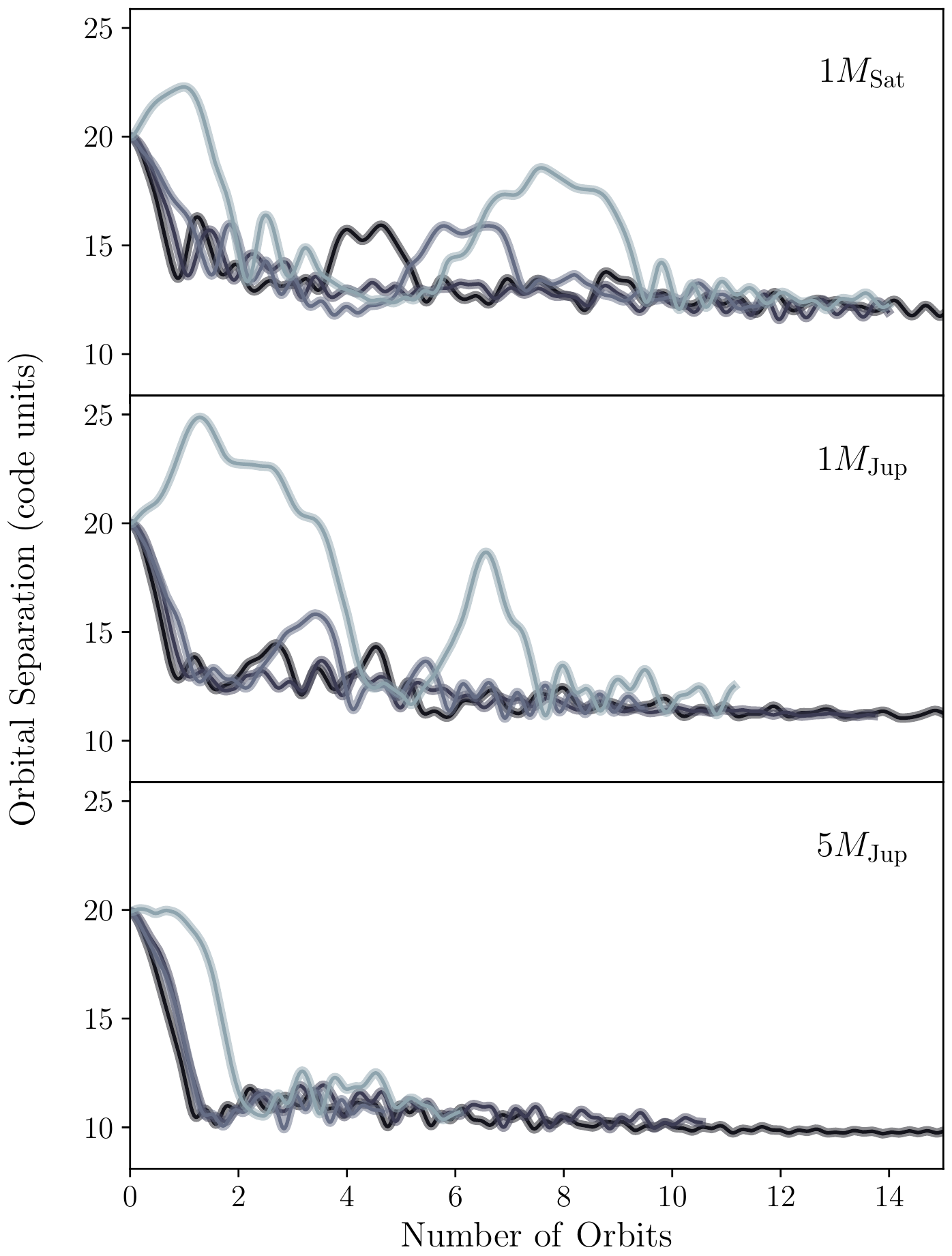}
\caption[]{Migration tracks of a $1M_{\mathrm{Sat}}$ (top), $1M_{\mathrm{Jup}}$ (middle), and a $5M_{\mathrm{Jup}}$ (right) planet started at different times to understand the effect of different turbulent environments. Despite the random turbulent structure around the planet at each start time, the planets always end up migrating inwards rapidly and slow down when they reach the gravitationally stable inner disc.}
\label{startTimes}
\end{center}
\end{figure}

The results from the various restarts are shown in Figure \ref{startTimes}. The migration of the giant planets are very similar to each other, regardless of when they start. In all cases, the initial rapid inward migration slows down in the gravitationally stable inner disc at roughly the same location and time. There is one exception, shown by the lightest line in Figure \ref{startTimes}, where the planets do not immediately migrate inwards. This is due to random chance, where the planet is {initially} embedded inside one of the spiral arms of the disc. In other words, the planet is in a region of much higher density fluctuation, increasing the stochasticity. This is seen in the migration tracks for all three planets, which fluctuates much more than the other runs. Nevertheless, even in this random scenario, the planet eventually slows down its migration in the inner disc.

\subsubsection{Low mass planets}

As seen in Figure \ref{startTimes} {(and as noted by \citealt{2011Baruteau})} the influence of stochastic kicks increases with decreasing planet mass. Hence, for low mass planets it is especially important to run multiple simulations as done so previously to observe the trend in the migration.  

\begin{figure*}
\begin{center}
\includegraphics[width= 0.85\linewidth]{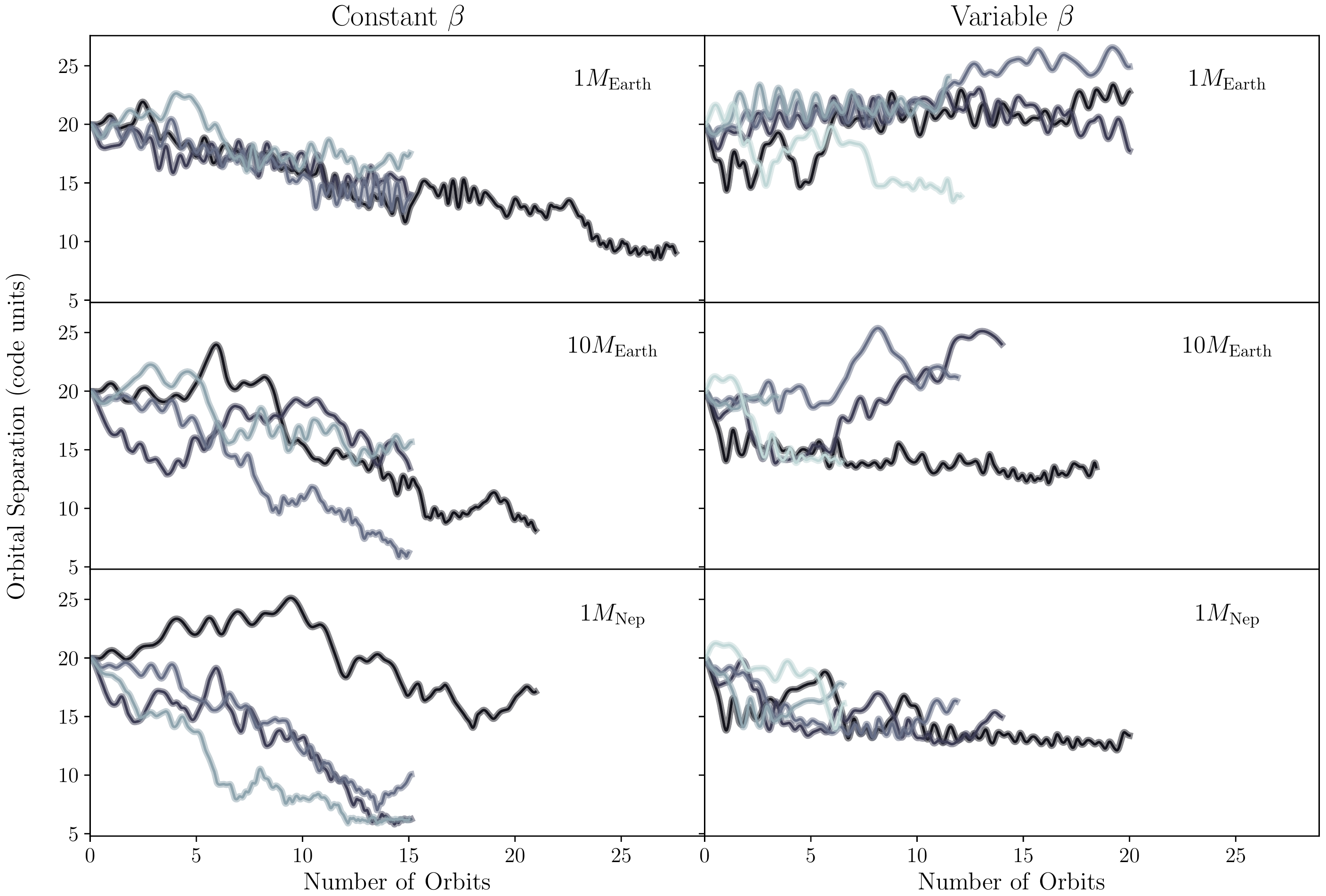}
\caption[]{Migration tracks of the $1M_{\oplus}$, $10M_{\oplus}$, and a $1M_{\mathrm{Nep}}$ planets started at different times for a constant (left) and a variable (right) $\beta$ disc. The general trend is that even low mass planets migrate to the inner edge of a constant $\beta$ disc. However, in a variable $\beta$ disc, like the giant planets, they slow down in the gravitationally stable inner disc.}
\label{LMPlanets}
\end{center}
\end{figure*}

We  begin by looking at the migration in the disc with a constant $\beta$, shown in the left column of Figure \ref{LMPlanets}. Although the low mass planets migrate inwards on a longer timescale compared to the giant planets, they show no signs of slowing down. A couple of simulations of the $10M_{\oplus}$  and $1M_{\mathrm{Nep}}$ planet even reach the inner edge. It is expected that with enough computational time, all simulations of the low mass planets would reach the inner edge.

As with the giant planets, the migration of the low mass planets in the disc with a variable $\beta$ (shown in the right column of Figure \ref{LMPlanets}) compared to a constant $\beta$ is quite different. In some cases, the $1M_{\oplus}$ and $10M_{\oplus}$ hardly show any inward migration. In the cases where the planets migrate inwards, the migration is significantly slowed down once the planet reaches the inner regions of the disc which is gravitationally stable. The slow down is best seen with the $1M_{\mathrm{Nep}}$ planet where after the initial rapid inward motion, the migration rate is much slower. With a variable $\beta$, there are some instances where the $1M_{\oplus}$ and $10M_{\oplus}$ planet do not migrate in at all. This is likely to be due to stochastic kicks, and will be discussed further in Section \ref{Torques}.

\section{Impact of numerics}
\label{sec:num}
\subsection{Resolution tests}

\begin{figure}
\begin{center}
\includegraphics[width= 0.85\linewidth]{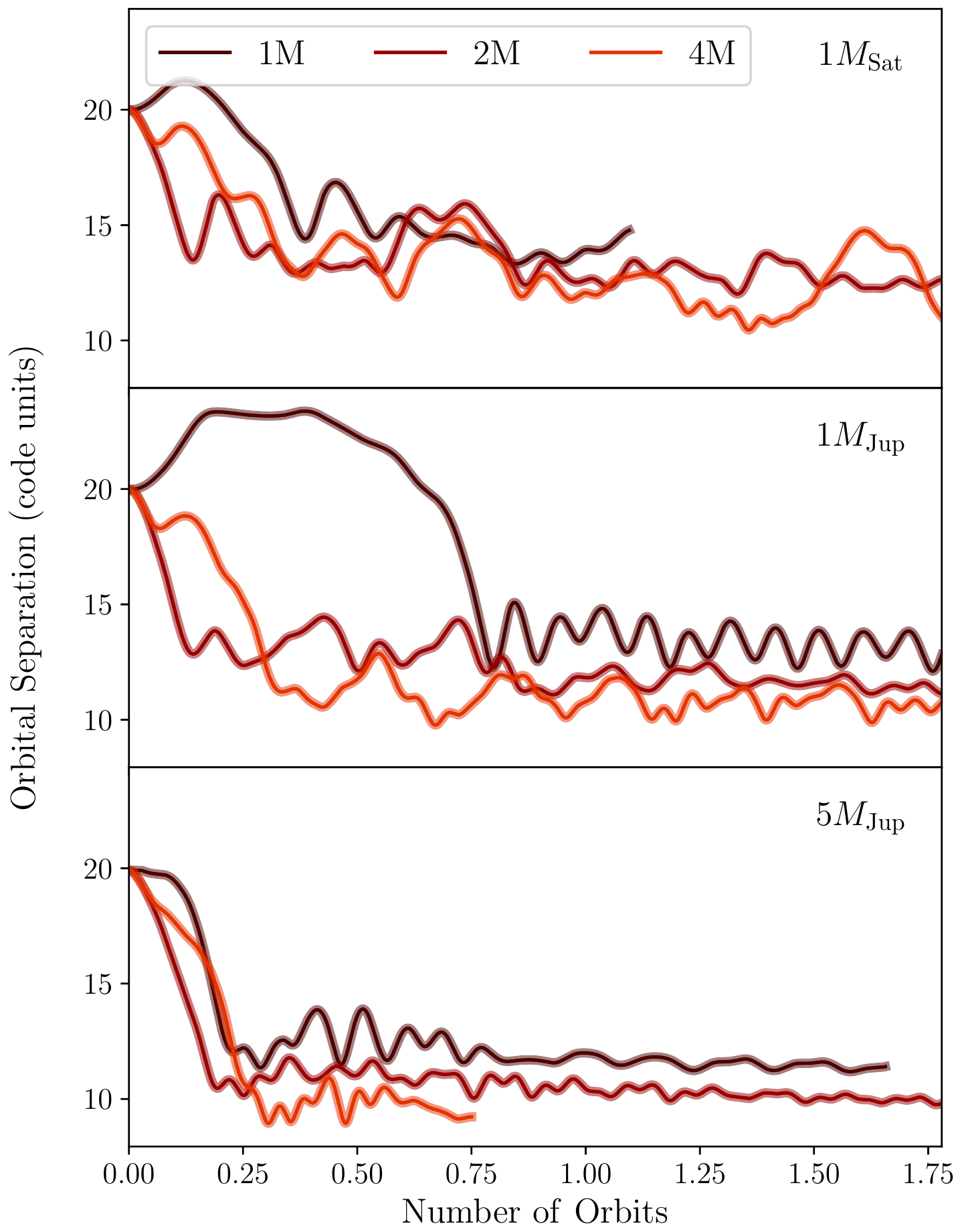}
\caption[]{Comparison of the planet migration in a disc with variable $\beta$ {modelled using 1, 2, and 4 million SPH particles}. The resolution increases with lighter shades. Inward migration slows down in the gravitationally stable inner disc in all cases.}
\label{res}
\end{center}
\end{figure}

To ensure that the above results are not affected by the resolution, we repeat the simulations with the giant planets, but with 1 and 4 million particles.  

The first check for any resolution effect is done by comparing the disc evolution. Both the low and high resolution broadly compared very well with the initial simulations; all 3 simulations are only gravitationally unstable in the outer regions of the disc.
{A minor difference is that the gravitationally unstable regime extends further  in with higher  resolution. However, the higher resolution also resolves more of the inner disc. Thus the extent of the gravitational stable region in all 3 discs is similar in size, but shifted further in with increasing resolution}. The planets are introduced at the same time in all three discs. The migration of each planet with different resolutions is shown in Figure \ref{res}.

\begin{figure}
\begin{center}
\includegraphics[width= 0.98\linewidth]{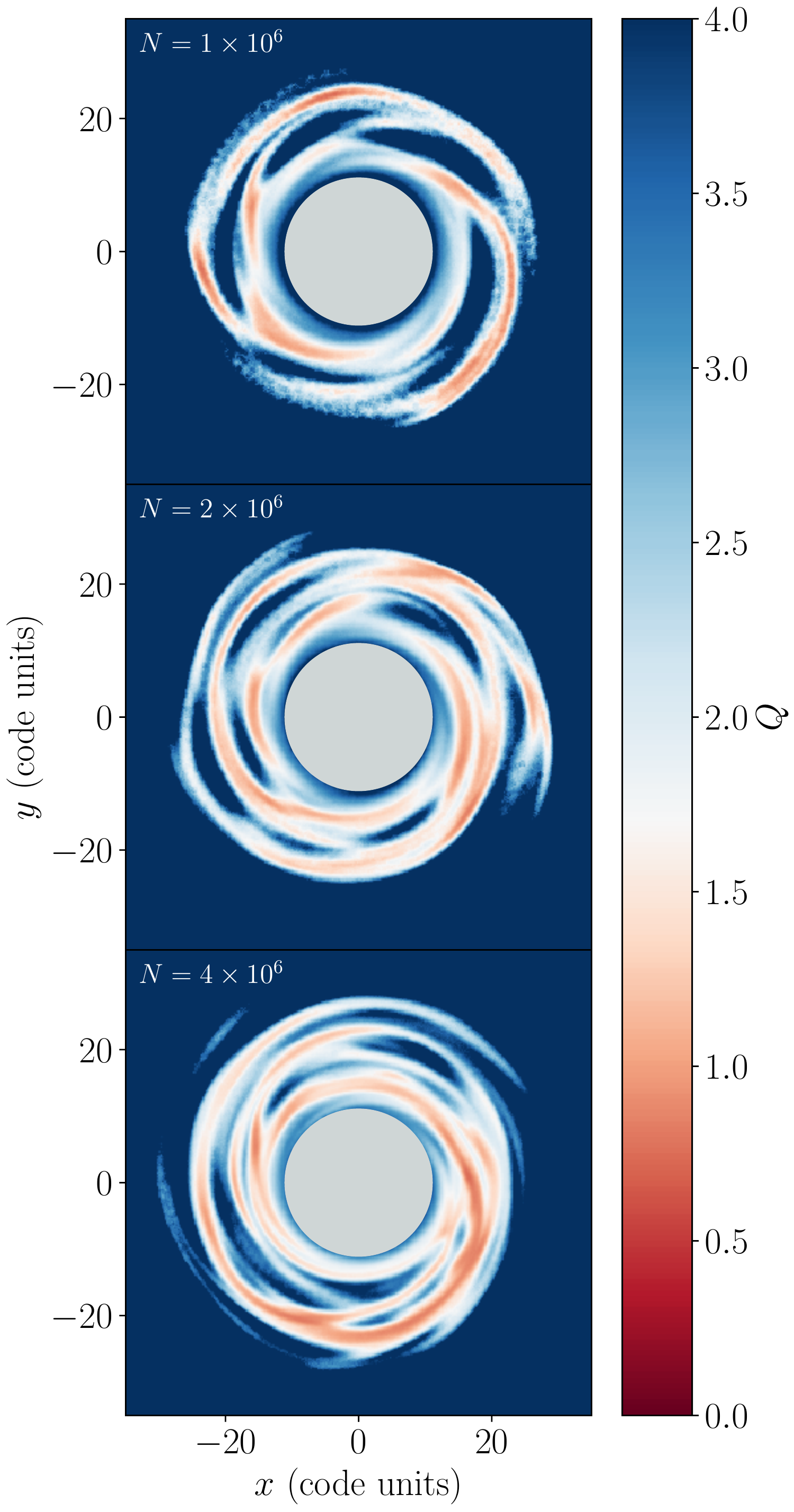}
\caption[]{2D maps of the Toomre parameter $Q$ of the variable $\beta$ disc just before the planets are added for a disc modelled with 1 million (top), 2 million (middle), and 4 million (bottom) particles. In red are the regions where the disc is gravitationally unstable. The size of the grey circle in each plot is the size of the gravitationally stable inner disc in the highest resolution simulation. It can be seen that in the lowest resolution simulation, more of the inner disc is gravitationally stable.}
\label{Q2dres}
\end{center}
\end{figure}

Although each planet does travel further in with increasing resolution, the conclusion that planet migration slows down in the inner gravitationally stable region of the disc remains unchanged. The further inward migration is due to the self-gravitating nature being resolved further in. This is shown in Figure \ref{Q2dres} where the 2D map of the Toomre parameter $Q$ is shown for all three resolutions just before the planet is added. In red are the regions which are gravitationally unstable. To compare how much of the inner disc is gravitationally stable, a {grey} circle is overlaid to represent the inner gravitationally stable disc of the 4 million particle simulation. From this it can be seen that as the resolution decreases, the gravitationally stable inner disc becomes larger. However, regardless of resolution, the migration  of a planet in a disc with variable $\beta$ always slows down in the gravitationally stable part of the disc and before reaching the inner disc edge.

\subsection{Different $\beta$ profiles}

Since the choice of how $\beta$ should vary with radius was arbitrarily chosen, we generalise it to 
\begin{equation}
\beta(R) = \beta_{0} \left ( \frac{R}{R_{0}} \right)^{-\delta}
\end{equation}
and investigate the impact of different values of  $\delta$. We repeat the previous simulations with $\beta = 27500 \ \left( R/R_{0} \right)^{-2.5}$ and ${\beta = 137500 \ \left( R/R_{0} \right)^{-3}}$, where $\beta_{0}$ is chosen such that ${\beta \left (R=25 \right) = 8.8}$ in all three cases. We do not investigate $\delta \leqslant 1.5$ since we require the cooling time, $t_{\mathrm{cool}}$ to decrease with radius so that only the outer regions become gravitationally unstable. With $\delta = 1.5$, the cooling time is constant throughout the disc.

A $1M_{\mathrm{Sat}}$, $1M_{\mathrm{Jup}}$, and $5M_{\mathrm{Jup}}$ planet is introduced at $R=20$ after the discs have evolved for 10 outer orbital periods. The different $\beta$ profiles do not change the key result as we find that the inward migration is slowed down when the planet reaches the inner gravitationally stable part of the disc.

\section{Discussion}
\label{sec:disc}

\subsection{Torque}
\label{Torques}

\begin{figure}
\begin{center}
\includegraphics[width= \linewidth]{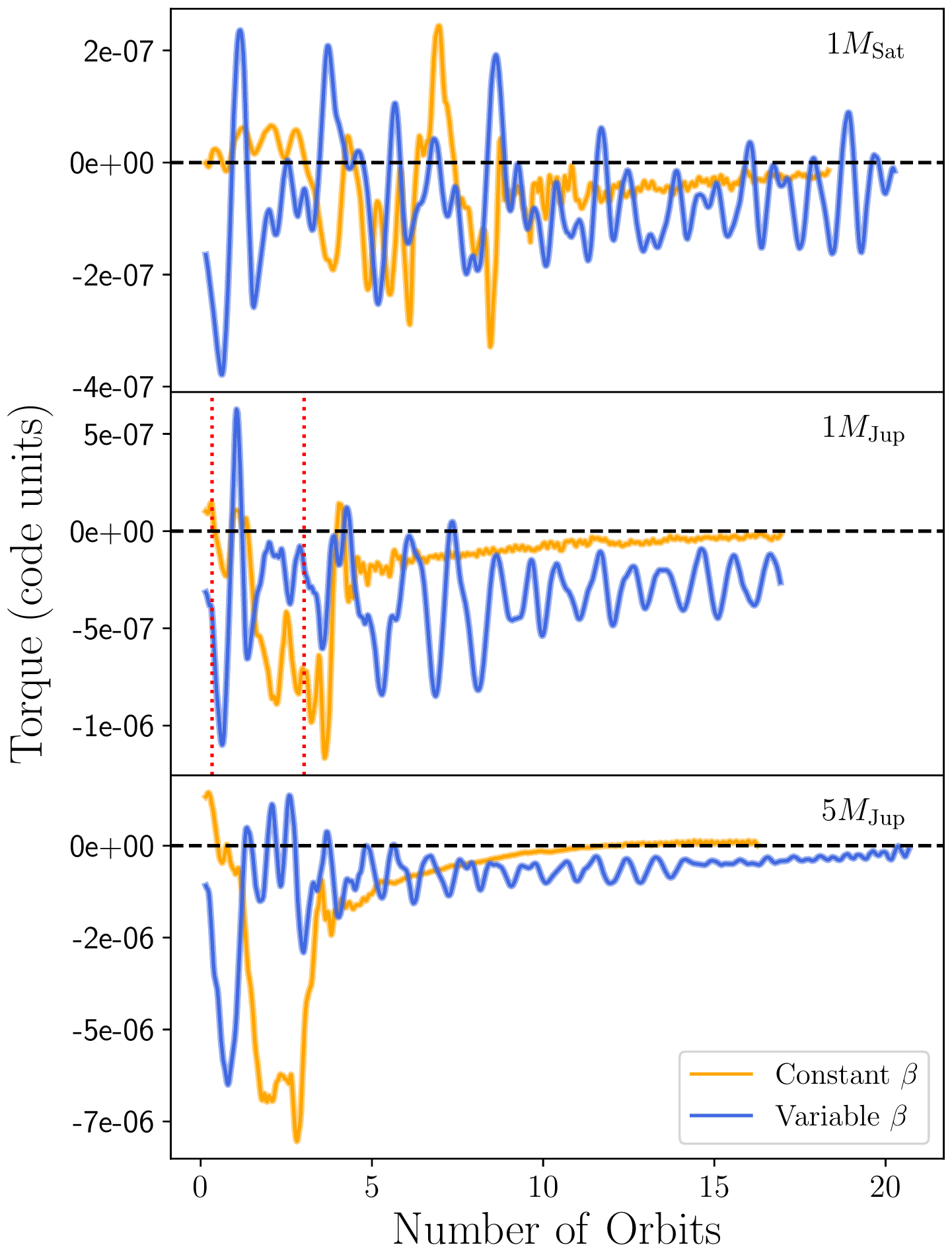}
\caption[]{Total $z$-component of the torque exerted by the disc material on a $1M_{\mathrm{Sat}}$ (top), a $1M_{\mathrm{Jup}}$ (middle), and a $5M_{\mathrm{Jup}}$ planet (bottom) in a variable $\beta$ (blue) and a constant $\beta$ (orange) disc. In a variable $\beta$ disc, the torque is most negative in the first orbit when the planet is migrating inwards rapidly. After which the torque is slightly negative, and thus the planet continues to migrate inwards but at a slower rate. In a constant $\beta$ disc, the torque is just as negative during the rapid inward migration, however it stays negative for longer and only decreases to near zero when the inner edge of the disc is reached. {The red dashed lines indicate the times shown in Figures \ref{TGPSigma} and \ref{aziT1J}.}}
\label{TGP}
\end{center}
\end{figure}

\begin{figure*}
\begin{center}
\includegraphics[width= 0.825\linewidth]{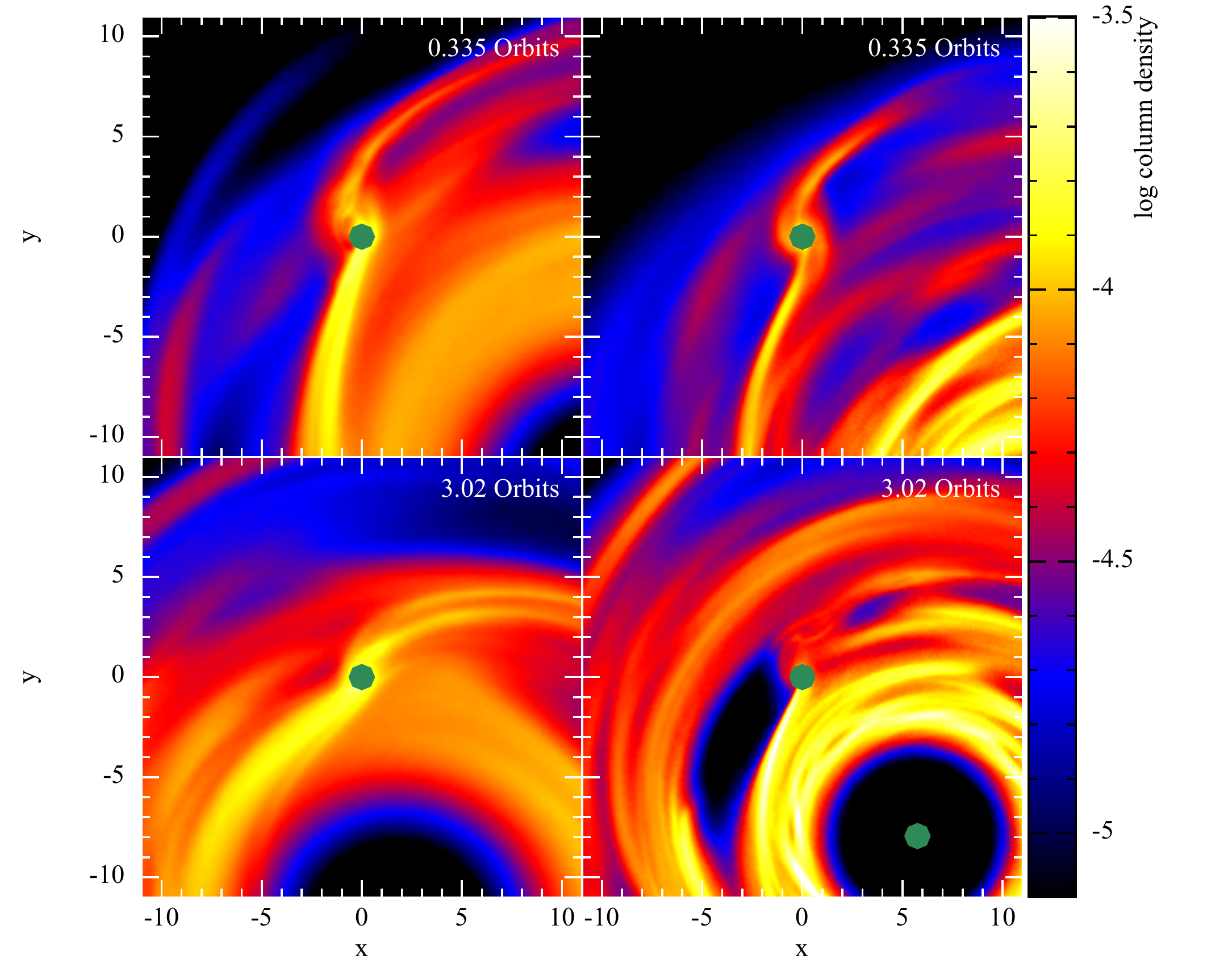}
\caption[]{This figure shows how the surface density of the disc near a $1M_{\mathrm{Jup}}$ planet changes as it migrates in a variable $\beta$ (left) and a constant $\beta$ (right) disc. Top left panel is when the planet is rapidly migrating inward in a variable $\beta$ disc. Bottom left is when the migration of the planet has slowed down. Top right is just before the planet starts rapidly migrating inwards in a constant $\beta$ disc. Bottom right is when the planet is rapidly migrating inwards. When the surface density of the region in front and behind the planet is comparable, there is little migration (bottom left and top right). However when there is an under-dense region in front and and over-dense region behind, the planet migrates inwards (top left and bottom right, also see middle panel of Figure \ref{migration}).}
\label{TGPSigma}
\end{center}
\end{figure*}

\begin{figure*}
\begin{center}
\includegraphics[width= 0.925\linewidth]{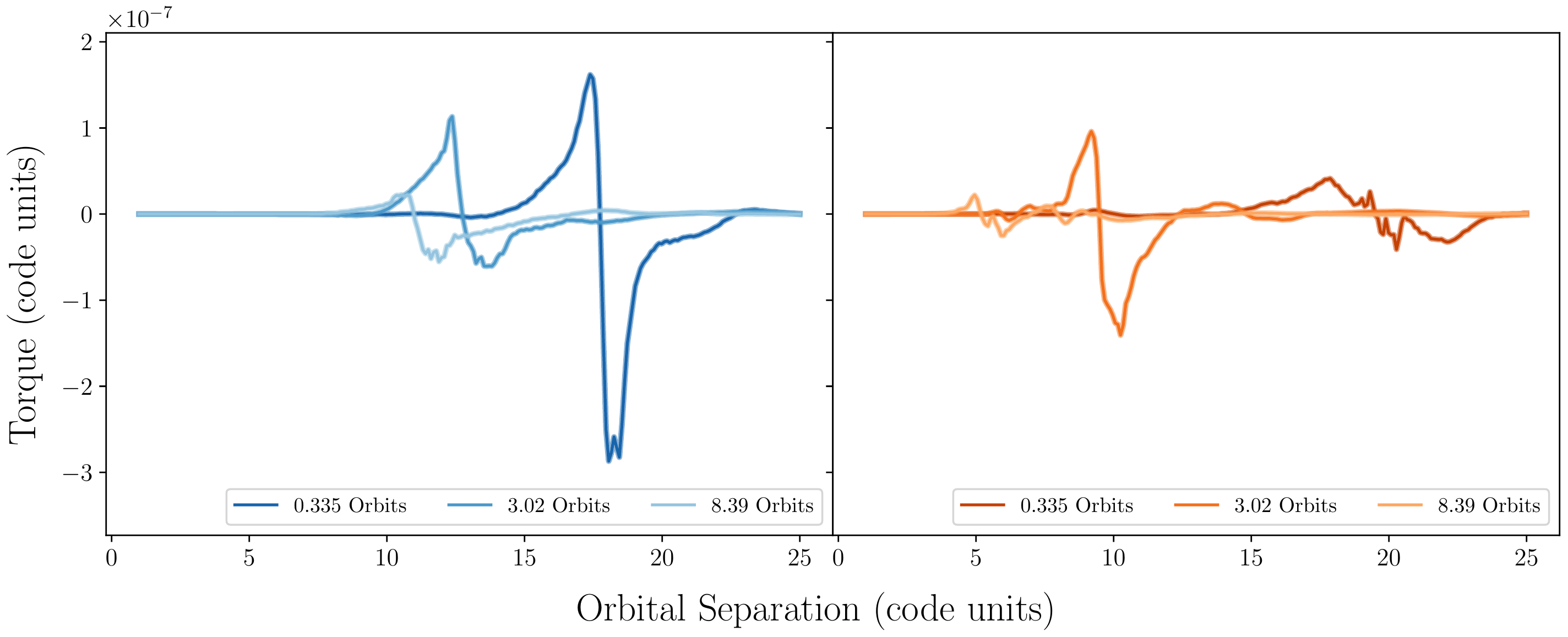}
\caption[]{Azimuthally-averaged torque acting on a $1M_{\mathrm{Jup}}$ as it migrates in a variable (left) and constant (right) $\beta$ disc. When the corotation torque is mostly negative, the planet is rapidly migrating inwards (compare times in Figure \ref{migrationPlot}) due to an underdense region in front of the planet (see Figure \ref{TGPSigma}).}
\label{aziT1J}
\end{center}
\end{figure*}

\begin{figure*}
\begin{center}
\includegraphics[width= \linewidth]{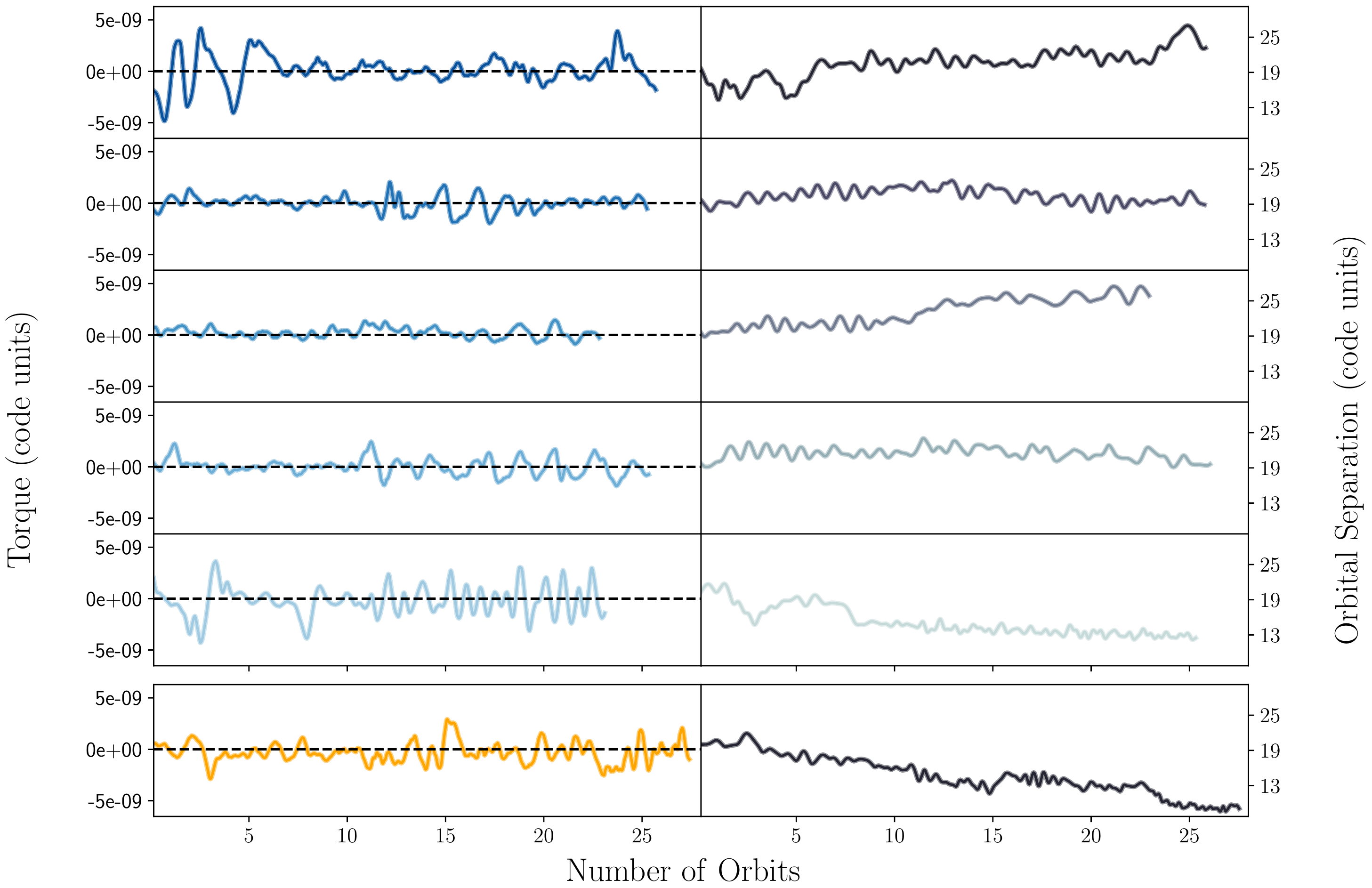}
\vspace{-0.4cm}
\caption[]{Total $z$-component of the torque exerted by the disc material on a $1M_{\oplus}$ planet is shown on the left panels. The migration tracks are shown on the right. The top 5 panels are simulations modelled with a variable $\beta$, whilst the bottom panel is one of the simulations modelled with a constant $\beta$. The migration is dominated by stochastic kicks which results in random outward or inward motion, as seen in the torques oscillating about zero.}
\label{T1E}
\end{center}
\end{figure*}

To investigate why the rapid inward migration of the planets slows down in a variable $\beta$ disc, we evaluate the tidal torque between the disc material and the planet.
The total $z$-component of the torque, which drives the radial motion of the planet, is calculated by summing the individual torque contributions from each fluid element on the planet and is given by
\begin{equation}
T_{z} = \sum_{i}^{N} \frac{G M_{\mathrm{p}} m_{i}}{d_{i}^{3}} \ \mathbf{r}_{\mathrm{p}} \times \mathbf{d}_{i}\ ,
\end{equation}
where $\mathbf{d}_{i}$ is the separation between each fluid element and the planet, $m_{i}$ is the mass of the fluid element, and $\mathbf{r}_{\mathrm{p}}$ and $M_{\mathrm{p}}$ are the position and mass of the planet respectively. To decrease numerical noise, {all material inside half the Hill radius of the planet is considered to be circumplanetary material and excluded from calculations. This exclusion zone varies with the stellar mass which increases as disc material is accreted by the star, and with the mass and location of the planet as it migrates}. The torques are also smoothed by taking the average over 0.2 orbits.

Figure \ref{TGP} compares how $T_{z}$ varies as the planet migrates in both a constant and variable $\beta$ disc. The behaviour of the torque is quite different in each disc. {With a constant $\beta$, the $1M_{\mathrm{Jup}}$ and $5M_{\mathrm{Jup}}$ planet experience a large negative torque until they reach the inner edge, at which point a lack of disc material results in the torque flattening to near zero very quickly. Due to random outward kicks while the  $1M_{\mathrm{Sat}}$ planet is migrating inwards, the torque profile is not continuously negative. Despite this, it is still clear that like the $1M_{\mathrm{Jup}}$ and $5M_{\mathrm{Jup}}$ planet, the magnitude of the torque only significantly decreases when it reaches the inner edge of the disc}. Whereas with a variable $\beta$, even though the torque once again quickly becomes just as negative, once the planet reaches the gravitationally stable inner disc, the magnitude of the torque decreases but remains negative resulting in more inward migration at a much slower pace. Unlike with a constant $\beta$, the torque does not flatten indicating it is still migrating inwards, albeit very slowly.

The difference in the torque evolution as the planet migrates can be explained by considering the disc structure near the planet as it migrates. The corotation torque, exerted by the gas within the horseshoe region plays a significant role in the magnitude and direction of the total torque. Asymmetries in the disc structure in this region can result in either inward or outward migration. Figure \ref{TGPSigma} shows the disc structure at important times during the planet's migration in a variable (left) and constant $\beta$ (right) disc. Comparing with Figure \ref{TGP} shows that the torque is most negative when the region in front of the planet is underdense relative to the region behind the planet (top left and bottom right). This density contrast results in a large negative co-oribital corotation torque which drives the planet inwards. This is seen in azimuthally-averaged torques in Figure \ref{aziT1J} which shows that when the planet is rapidly migrating inwards, the corotation torque is mostly negative. Whereas when there is little density contrast around the planet (top right and bottom left in Fig \ref{TGPSigma}), the torque profile in Fig \ref{aziT1J} is more symmetric resulting in slower migration.

As expected, comparing the torque on the  $1M_{\mathrm{Sat}}$ to the $1M_{\mathrm{Jup}}$ and $5M_{\mathrm{Jup}}$ planets show that lower mass planets are more affected by stochastic kicks. Unlike the latter two, the torque on a $1M_{\mathrm{Sat}}$ frequently becomes positive resulting in the sporadic moments of small outward migration as seen in Figure \ref{migration}. However, it is still clear the torque on average is negative and the planet is still migrating inwards. 

However, the increased effect of stochastic kicks as planet mass decreases could explain why some of the low mass planets do not migrate inwards. Figure \ref{T1E} shows the migration (right column) of a $1M_{\oplus}$ planet, and how $T_{z}$ (left column) varies as it migrates. The first five rows are the different runs in a variable $\beta$ disc. In nearly all cases, the migration is dominated by stochastic kicks as seen by the torque fluctuating about zero, resulting in small inward and outward migration. The bottom row is in a constant $\beta$ disc, where it is clear that despite stochastic kicks being present, the torque on average is slightly negative resulting in inward migration.

\subsection{Comparison with previous work}

We compare our results to those of \cite{2011Baruteau} and \cite{2015Malik} who considered giant planet migration in self-gravitating discs. There are a couple of differences in this work. They perform their simulations using a 2D grid based code, whereas the simulations in this work are performed using a 3D SPH code. The initial conditions of the disc are also slightly different. {Their disc surface density and temperature decreased as $R^{-2}$ and $R^{-1}$ respectively, whereas in this work, the disc surface density and temperature decreases as $R^{-1}$ and $R^{-0.5}$ respectively}. To ensure that the different initial conditions and codes used do not influence the results, we initially	 performed a  set of simulations with a constant value of $\beta = 15$ to compare with \cite{2011Baruteau} and \cite{2015Malik}. Despite the differences, we agree with their findings that using a constant $\beta$ results in the giant planets rapidly migrating towards the inner edge of the disc.

{However, the crucial development in our work is how the disc is cooled: whilst, we also utilise the $\beta$-cooling approach, we allow $\beta$ to vary with radius to mimic a more realistic disc such that our discs remain gravitationally stable in the inner regions, and spiral structure only forms in the outer regions. The planets are able to slow down their rapid inward migration before reaching the inner disc edge.	}

Using radiative transfer, \cite{2018Stamatellos} migrate $1M_{\mathrm{Jup}}$ planets in self-gravitating discs {using a 3D SPH code}. The initial conditions in their discs are more similar to ours; the same initial disc mass and surface density profile. Their usage of a realistic treatment of the disc thermodynamics meant that like in the simulations presented here, the planets are evolved in discs that are only gravitationally unstable in the outer regions. Using realistic thermodynamics to allow the gas to control its cooling/heating based on its properties, the thermodynamics of the circumstellar and circumplanetary disc is better captured. This enhances the mass growth of the planet causing the planets to grow to beyond the brown dwarf limit. This resulted in their planets opening up gaps and slowing their inward migration. Thus it is unclear whether the mass growth or its presence in the more gravitationally stable inner disc was the dominant factor in slowing the inward migration. {We do not allow our planets to grow and hence show that modelling the cooling in the discs such that only the outer disc is gravitationally unstable, enables a planet to slow its migration in the inner gravitationally stable disc. Therefore, while we expect the growth of a planet would help slow its migration, our work shows that {the location of the planet in the disc} has an important effect too.}

{Recently \cite{2018Mercer}, using their radiative transfer simultions calculated an effective value of beta, $\beta_{\mathrm{eff}}$. Using this, they showed that a constant $\beta$ does not accurately reflect a realistic gravitationally unstable protoplanetary disc. Their calculated $\beta_{\mathrm{eff}}$ was found to vary both spatially and with time as the disc evolves. Since the variable $\beta$ model in this work does not vary with time, to achieve a similar comparison, we compare the variable $\beta$ profile here with their $\beta_{\mathrm{eff}}$ of a high mass disc that has not yet developed spiral arms. In general, the two compare reasonably well with both decreasing as roughly $R^{-2}$ in the inner regions of the disc. The main difference between the two is in the outer regions where  the $\beta_{\mathrm{eff}}$ of \cite{2018Mercer} decreased more slowly tending towards a non-zero $\beta$. However this difference could be enhanced by the differences in the simulation setup, where their simulations have a ambient radiative field of $10\mathrm{K}$ and they simulate a higher disc-to-star mass ratio. Furthermore, their discs eventually fragment, whereas the $\beta$ profile chosen here is such that disc fragmentation is avoided.}

\subsection{In the context of observations}

The viability of early planet formation is important given recent observations of young protoplanetary discs. Some of these discs are a few million years old or younger \citep{2018Andrews} and already show substructure such as rings and gaps.   Assuming that the gaps were carved out by planets, \cite{2019Lodato} determined that a wide range of planet masses from super Earths to a few Jupiters at wide orbits ($10-100$ AU) were typically required. There is also an increasing body of evidence  \citep{2018Clarke, 2019Flagg} that show that planets have formed in young discs ($\sim 1$Myr). 

Although the discs in the aforementioned studies may not be expected to be gravitationally unstable given their mass estimates, they still indicate that planets can form very early in a disc's lifetime. However, measuring the disc mass can be quite challenging. Recently \cite{2020Booth} using the rarer $\mathrm{CO}$ isotopologue, $^{13}\mathrm{C}^{17}\mathrm{O}$ measured the mass of HL Tau to be at least $0.2M_{\odot}$, up from the previous gas mass measurements of $2-40 \times 10^{-3}M_{\odot}$ \citep{2018Wu}. This puts HL Tau in the gravitationally unstable regime beyond 50AU, where a potential planet might be carving the gaps seen. 

{If planets can form in young discs, the next most natural disc to ask is whether they can survive their migration}. Our work shows that planets formed -- by any process -- in young discs can survive and evolve in the disc. 

\subsection{In the context of population synthesis models}

The work presented here is also important for studying population synthesis models in self-gravitating discs, such as the ones in \cite{2013Galvagni}, \cite{2013Forgan}, and \cite{2018Forgan}. A common problem in these models is using the rapid migration times. This results in very low survivability of any clumps formed via gravitationally stability, unless the clumps could open up a gap to slow their migration and survive.  The results here show that with more realistically modelled thermodynamics, the inward migration of planets in gravitationally unstable discs can be suppressed without requiring gap-opening. Further work will be required to study the impact that a slow down of a planet's migration  has on population synthesis studies.

\subsection{Caveats}

The main goal of this paper is to study the impact of implement a more realistic treatment of the thermodynamics on planet migration. Thus, the main difference compared to past studies with {$\beta$-cooling} models is how $\beta$ varies with radius. Several other possible important parameters are kept fixed.

We only consider one disc model. The temperature and surface density profiles are the same for all simulations.   The disc mass is also kept fixed. Although these would change how the Toomre parameter evolves, and thus how the gravitational stability of the disc evolves, it is not expected to change the result that planet migration can be slowed down in the gravitationally stable inner disc.

We have also {limited} accretion onto the planets. Studies have shown that disc fragments (${\sim} 1-10 M_{J}$ \citealt{2010Boley}) can easily accrete enough material to become brown dwarfs \citep{2009Stamatellos,2010Kratter,2012Zhu}. To keep things simple, the accretion radii of the planets are small enough to ensure the planet's mass stays {roughly} constant. This is obviously not very realistic, but is done firstly to directly compare with \cite{2011Baruteau} and \cite{2015Malik} and secondly to understand the impact that different locations of gravitationally unstable discs have on planet migration and survivability.

Despite using a variable $\beta$ to mimic the varying cooling in a real self-gravitating disc, the cooling in a disc is likely to be much more complex. Further work will be done in the future comparing the results here with a disc modelled using radiative transfer.

\section{Conclusion}
\label{sec:conc}

We perform 3D SPH simulations to investigate the migration of both giant planets with masses of $1M_{\mathrm{Sat}}$,  $1M_{\mathrm{Jup}}$, and  $5M_{\mathrm{Jup}}$, and low mass planets with masses of $1M_{\mathrm{\oplus}}$,  $10M_{\mathrm{\oplus}}$, and $1M_{\mathrm{Nep}}$  in self-gravitating discs. Our study shows that by implementing a radially decreasing cooling time to more accurately cool the disc such that only the outer region of the disc is gravitationally unstable, the inward migration of planets is slowed down in the inner gravitationally stable part of the disc {without requiring a gap to open up}. These results show, most importantly that planets may be able to survive their inwards migration in real self-gravitating discs.

\section*{Acknowledgements}

We thank Cathie Clarke, Colin McNally, and Ken Rice for useful discussions which benefited this work. SR acknowledges support from the Royal Society Enhancement Award. We also thank the anonymous referee for their useful comments. FM acknowledges support from the Royal Society Dorothy Hodgkin Fellowship.  This work has made use of the SPLASH software \cite{2007Price}. This work was performed	using Orac and Tinis HPC clusters at the University of Warwick.



\bibliographystyle{mnras}
\bibliography{PlanetMigration}


\bsp	
\label{lastpage}
\end{document}